\documentclass[aps,prl,twocolumn,groupedaddress,nofootinbib]{revtex4}
\usepackage[dvips]{graphicx}
\usepackage{color}
\usepackage{amsmath,amssymb,slashed}
\newif\ifdraft
\drafttrue
\newif\ifpreprint
\preprinttrue

\def\spa#1.#2{\left\langle#1\,#2\right\rangle}
\def\spb#1.#2{\left[#1\,#2\right]}

\newcommand{\eq}{\begin{equation}}
\newcommand{\eqe}{\end{equation}}
\newcommand{\eqa}{\begin{eqnarray}}
\newcommand{\eqae}{\end{eqnarray}}

\newcommand{\bea}{\begin{eqnarray}}
\newcommand{\eea}{\end{eqnarray}}

\newbox\charbox
\newbox\slabox
\def\s#1{{      
        \setbox\charbox=\hbox{$#1$}
        \setbox\slabox=\hbox{$/$}
        \dimen\charbox=\ht\slabox
        \advance\dimen\charbox by -\dp\slabox
        \advance\dimen\charbox by -\ht\charbox
        \advance\dimen\charbox by \dp\charbox
        \divide\dimen\charbox by 2
        \raise-\dimen\charbox\hbox to \wd\charbox{\hss/\hss}
        \llap{$#1$}
}}

\begin{document}
\title{
\begin{flushright} ${}$\\[-40pt] $\scriptstyle \rm NCTS-TH/1610$ \\[0pt]
\end{flushright}
On the linearity of Regge trajectory at large transfer energy} 

\author{
Carlos Cardona$^{a}$, Yu-tin Huang$^{a,b}$,
and Tsung-Hsuan Tsai $^{b}$}

\affiliation{$^a$ Physics Division, National Center for Theoretical Sciences, National Tsing-Hua University,
No.101, Section 2, Kuang-Fu Road, Hsinchu, Taiwan}
\affiliation{$^b$ Department of Physics and Astronomy, National Taiwan University, Taipei 10617, Taiwan}

\begin{abstract}
In this note we consider the four-point function of identical scalar operators in its Mellin representation. For CFTs, taking the large scaling dimension limit of the Mellin amplitude yields the flat-space S-matrix. Using this correspondence, and assuming that the Regge-limit of the Mellin amplitude vanishes, we prove that the resulting flat space S-matrix must have string-like linear Regge trajectory in the limit $|s|>>|t|>>1$. We also show that the Regge limit of Mellin amplitude is captured by the three-point coefficient of two scalars and one double trace operator in the large-spin limit. Using this, we can see that vanishing Regge behaviour occurs at small t'Hooft limit. 
\end{abstract}

\maketitle

\section{Introduction}  
The constraint of unitarity, locality and causality take their sharpest form on the S-matrix, and can impose non-trivial restriction on the infrared (IR) effective field theory, or the ultra-violet (UV) completion itself. Notable examples are the positivity constraints on coefficients of higher-dimensional operators in the IR, derived from the assumption of an UV S-matrix satisfying the usual analyticity constraints~\cite{Adams:2006sv}. For examples of constraints on the UV completion itself, one can point to that of causality~\cite{Juan}, which shows that if the gravitational three-point coupling deviates from the Einstein-Hilbert interactions while the theory is weakly coupled, the usual Shapiro-Time delay develops advancement, and hence violates causality. This violation can only be cured with the inclusion of a infinite tower of massive higher spin states.  

Assuming that a theory is UV completed by an infinite tower of massive higher spin states, it would be interesting to understand what constraints on the spectrum does unitarity imposes. A priori, one might expect that there are no constraints, since one can easily construct a four-point amplitude with arbitrary massive higher-spin exchange that factorizes consistently. This would be a function of two variables $M(s,t)$ where $s=(p_1+p_2)^2$ and $t=(p_1+p_4)^2$, and unitarity simply tells us that the function can be written as a sum over simple poles with its residue given by a positive coefficient multiplying the Gegenbauer polynomials. 

While this certainly does not constrain the spectrum, non-trivial results can be obtained if we further assume certain high energy behaviour, such as the polynomial boundedness of the amplitude at large $|s|$. In this limit $s>>1$ with fixed $t$, referred to as the Regge limit, the amplitude generically takes the asymptotic form $s^{j(t)}$, where the function $j(t)$ is the Regge trajectory.  The solution to $j(t)=n$, with $n$ being positive integers gives the spectrum of the theory.  This limit was recently carefully studied in~\cite{Caron-Huot:2016icg}, and using properties of positive functions, it was shown that in the asymptotic Regge limit, $s>>t>>1$, the Regge trajectory is linear, i.e. $j(t)=\alpha' t+\mathcal{O}(1)$.

On the other hand the Mellin amplitude, which appears in the Mellin transform of correlation functions, have similar analytic structures as a physical scattering amplitude. Assuming a perturbative expansion for the anomalous dimensions (of double trace operators), it is then a meromorphic function of simple poles, and unitarity of the correlation function can be directly translated into unitarity of the Mellin amplitude~\cite{Mack, Raju}. In AdS/CFT, taking the large scaling-dimension limit on the CFT side then correspond to the flat space limit of AdS Witten diagrams. This relation has been successfully utilised to obtain constraints on anomalous dimensions~\cite{Alday:2016htq}, as well as obtain bounds on cubic couplings for the flat-space S-matrix~\cite{Pedro, Pedro1}.

In this letter we will use this correspondence to study scenarios where the Mellin amplitude vanishes in the Regge limit. This implies that saddle point must exist in the large $s$ limit for particular value of $t$. On the other hand, taking the limit where the two cross-ratios asymptote to $1$ and $0$ respectively (position space Regge-limit), precisely probe the Regge limit for the Mellin amplitude~\cite{CRegge}. 
We will show that in taking the flat space limit, the position space Regge-limit defined above actually probes the absolute asymptotic Regge limit, $|s|>>|t|>>1$. Studying in the Mellin integral in this limit, and requiring that a saddle point indeed exists for some value of $t$, we will show that vanishing Regge limit implies a linear trajectory, in the limit $|s|>>|t|>>1$.

\vspace{-0.3cm}

\vspace{-0.4cm}

\section{The Mellin Amplitudes}
We begin with the reduced four-point correlator $\mathcal{A}(u,v)$ for four identical scalar operators. This is defined through the four-point function as,
\eq
\langle\mathcal{O}(x_1)\mathcal{O}(x_2)\mathcal{O}(x_3)\mathcal{O}(x_4)\rangle=
\frac{1}{(x_{12}^2)^{\Delta} (x_{34}^2)^{\Delta}}\mathcal{A}(u,v) \,,
\eqe
where $\Delta$ is the dimension of the scalar operator, and $u,v$ are the two independent cross ratios, 
\eq
u= \frac{x_{12}^2 x_{34}^2}{x_{13}^2 x_{24}^2}\ ,\ \ \ \ \ \ \ 
v= \frac{x_{14}^2 x_{23}^2}{x_{13}^2 x_{24}^2}\ .
\eqe
Through a Mellin transform, the reduced correlator is expressed as: 
\eqa\label{Def1}
\mathcal{A}(u,v)&=&\int_{{-}i\infty}^{i\infty}\frac{dtds}{(4\pi i)^2}M(s,t)\frac{u^{\frac{t}{2}}}{v^{\frac{s}{2}-\Delta}}\Gamma\left(\frac{t{+}s}{2}-\Delta\right)\nonumber\\
&&\Gamma\left(\Delta{-}\frac{t}{2}\right) \Gamma\left(\Delta{-}\frac{s}{2}\right)\,,
\eqae
where $M(s,t)$ is the Mellin amplitude, which is a meromorphic function with poles in $s,t$. The integral in eq.(\ref{Def1}) is understood as a contour integral, picking up poles from the the gamma functions as well as that in the Mellin amplitude.\footnote{ The choice of Gamma functions are such that the canonical dimensions of ${\footnotesize \mathcal{O}(\overleftrightarrow{\partial})^{2n}\mathcal{O}}$ are correctly reproduced.}

The position of the poles in $M(s,t)$ can be read from the OPE expansion of the four-point function. The reduced four-point function can be represented as
the conformal block expansion of the reduced correlator given by
\eq\label{CBE}
\mathcal{A}(u,v)= \sum_k C_{12k}C_{34k} \, G_{\Delta_k,J_k}(u,v)\,.
\eqe
where one sums over the primary operators in the theory, labeled by its dimension and spin $(\Delta_k,J_k)$, while $G_{\Delta_k,J_k}(u,v)$ are the conformal blocks. In the limit $u\rightarrow0, v\rightarrow1$ the dependence on $u$ in eq.(\ref{CBE}) is inherited from the limiting form $G_{\Delta_k,J_k}(u,v)|_{u\rightarrow0}\sim u^{\frac{\Delta_k-J_k}{2}}$. To reproduce this behaviour, one can deduce that $M(s,t)$ must have poles at $t=\Delta_k-J_k+2m$, where $m=0,1,2,\cdots$ represents the descendants. 

On the other hand, using the crossing symmetry equation for the four-point function,
\eqa\label{crossing}
\mathcal{A}(u,v)=\left(\frac{u}{v}\right)^\Delta \mathcal{A}(v,u)
\eqae 
and substituting the Mellin representation in eq.(\ref{Def1}), one finds that in order for eq.(\ref{crossing}) to hold, the Mellin amplitude must be symmetric: $M(s,t)=M(t,s)$.
\section{The Flat Space Limit of Mellin Amplitudes}
Mellin amplitude has the same analytic property as that of scattering amplitudes of massive particles mass $\Delta$ and poles in the $t,s$-channel. To make connection with tree-level amplitudes, one must assume that the CFT has an expansion parameter such that the anomalous dimension of the composite operator ${\footnotesize \mathcal{O}(\overleftrightarrow{\partial})^{2n}\mathcal{O}}$ is perturbative. The reason is that the exchange of such operators can be thought of as exchanging two particles in a loop diagram. We wish to separate this part of the Mellin amplitude, such that only the tree-level piece is retained. As the classical part of the ${\footnotesize \mathcal{O}(\overleftrightarrow{\partial})^{2n}\mathcal{O}}$ exchange is given by the poles of the Gamma functions in eq.(\ref{Def1}), this is achieved if the anomalous dimension can be perturbatively separated. In the usual context of AdS/CFT, this correspond to the large $N$ parameter.

Now let's assume that for some fixed value of $t=t^*$, the Mellin amplitude vanishes in the Regge limit 
\eq\label{ReggeS}
M(s,t^*)_{|s|\rightarrow\infty}=0\,.
\eqe
If this is the case, then in the large $s$ limit, with $s$ slightly off the real line, while the Mellin amplitude is decreasing, the Gamma functions in eq.(\ref{Def1}) are growing. This implies that there a saddle point must exist. In the next section, we will use this as a consistency condition on the position space Regge-limit of the reduced correlation function.

Before moving on, it is interesting to ask if one can identify CFTs whose Mellin amplitude exhibit such Regge asymptotic. Consider the contribution of pole in the Gamma functions for $s=2\Delta$, and $t=2\Delta+2n$, which is proportional to 
\eq\label{Residue}
M(2\Delta,2\Delta+2n)u^{\Delta+n}\frac{\Gamma[\Delta+n]}{n!}\,.
\eqe
Using $M(s,t)=M(t,s)$, taking $n>>\Delta$ then probes the Regge limit of the Mellin amplitude. By comparing with the $u\rightarrow0, v\rightarrow1$ limit of the expansion eq.(\ref{CBE}) where the conformal blocks can be written in terms of Gegenbauer polynomials (see \cite{Dolan:2011dv}), we see that for finite $n$, eq.(\ref{Residue}) corresponds to exchanges of ${\footnotesize \mathcal{O}(\overleftrightarrow{\partial})^{2n'}\mathcal{O}}$ operators, with $n'\geq n$. Thus taking the $n>>\Delta$ then isolates the exchange of a large spin operator. If the Mellin amplitude behaving as $s^a$ in the Regge limit, this then translate to the large-spin limit of the OPE coefficient behaving as:
\eq\label{3pt}
C_{\mathcal{O}\mathcal{O}{\footnotesize (\mathcal{O}\overleftrightarrow{\partial}^{2n}\mathcal{O}})}\sim n^{\frac{\Delta+a-1}{2}}\,.
\eqe  
Thus studying its dependence on spin yields information on the Regge-limit of the Mellin amplitude. Comparing with free field theory, where the coefficient is given by~\cite{Heemskerk:2009pn}:\footnote{This is valid for $D=2,4$, but is believed to be valid for arbitrary even dimensions. We thank L. Alday for pointing this out.}
\eq
C^{free}_{\mathcal{O}\mathcal{O}{\footnotesize (\mathcal{O}\overleftrightarrow{\partial}^{2n}\mathcal{O}})}=\left(\frac{\Gamma^2[\Delta+2n]\Gamma[2\Delta+2n-1]}{2n!\Gamma^2[\Delta]\Gamma[2\Delta+4n-1]}\right)^\frac{1}{2}\,
\eqe
which in the large $n$ limit becomes $n^{\Delta-\frac{3}{2}}$.  Matching with eq.(\ref{3pt}) we see that $a=\Delta-2$, which is consistent with the linear stringy Regge trajectory, since in that case we will expect $a\sim  t\sim \Delta $. Thus analytically continuing to small or negative $\Delta$, we obtain vanishing Regge limit. Thus vanishing Regge-limit correspond to small t'Hooft coupling for large $N$ theories.

At this point, one may be tempted to utilise the connection between Mellin amplitudes and reduced correlator to study general features of amplitudes with vanishing asymptotics. However, due to the presence of descendants, the positions of the poles are at $\Delta-J+m$, and this would appear that each particle must come with an infinite number of neighbouring states with mass differing by integer multiples of some fundamental scale. This is a rather restricted spectrum and is in no sense general from the view point of the scattering amplitude. 

Instead, we now take the flat space limit, which corresponds to taking $\Delta_i-$large, while keeping the ratio of $\Delta_i/\Delta_j$ fixed. \footnote{The terminology has its origin from holography, where the dimension of the operator is related to the mass of the bulk state as $\Delta(\Delta-d)=R^2m^2$, and hence large $\Delta$ can be viewed as large AdS radius $R$. However this analysis does not require holographic principle.} As discussed in~\cite{Pedro}, under this limit, the sum over descendants can be replaced by an integral whose value is peaked around a particular $m^*$ ( for scalar exchanges $m^*=(2\Delta-\Delta_k)^2/8\Delta$), with a width of order $\mathcal{O}(\sqrt{\Delta})$. Thus in the $\Delta_i-$large limit, the sum over descendants is replaced with a single pole in the amplitude. Thus the spectrum of the massive states in the flat space S-matrix is identified with the spectrum of primary operators in the CFT.

\section{The asymptotic Regge limit}\label{Sec2}
In \cite{CRegge}, it was shown that the $v\rightarrow1$ limit in spacetime correspond to Regge limit of the Mellin amplitude.\footnote{In \cite{CRegge} the parameterisation $u=\sigma^2$ and $v=1-2\sigma\cosh \rho$ is used, and hence $v\rightarrow1$ limit entails $u\rightarrow0$. However one can easily imagine other configurations such as $v=u=1$, which correspond to the four operators sitting on the vertices of a simplex. } This was done by demonstrating that in the large imaginary $s$ region of the Mellin integral develops a large oscillating phase that is minimised when $v\rightarrow1$. Taking the flat space limit, following the same argument we will show that $v\rightarrow1, u\rightarrow0$ correspond to the asymptotic Regge limit.

We take the flat space limit as $\Delta\sim |s|\sim |t| >>1$, where the absolute value is due to the fact that the Mellin integral is along the imaginary axes. All of the Gamma functions inside the integration eq.\eqref{Def1} can then be approximated by the Stirling formula eq.\eqref{Stirling},
\eq\label{Stirling}
\Gamma(z)\approx\sqrt{2\pi}e^{-z}(z)^{z-1/2} ,\,\quad |\rm{Arg}(z)|<\pi ,\,~~ |z|>>1\,.
\eqe
leaving us with,
\eqa
&&\mathcal{A}(u,v)\approx\nonumber\\
&&(2\pi)^{\frac{3}{2}}\int^{i\infty}\frac{dtds}{(4\pi i)^2}\widetilde{M}(s,t)\frac{u^{\frac{t}{2}}}{v^{\frac{s}{2}-\Delta}e^{i\pi s}}\left(\frac{t+s}{2}-\Delta\right)^{\frac{t+s}{2}-\Delta}\nonumber\\
&&e^{-\Delta}\,\left(\Delta-\frac{s}{2}\right)^{\Delta-\frac{s}{2}}\left(\Delta-\frac{t}{2}\right)^{\Delta-\frac{t}{2}}\,,
\eqae
where we have denoted $\widetilde{M}(s,t)$ as the flat space limit of the Mellin amplitude.\footnote{There is a subtlety with regards to the preferred causal relations in the limit, with $x^2_{14}x^2_{23}<0$ and $x^2_{12}x^2_{34}>0$. This requires us to rotate $v\rightarrow e^{i2\pi}v$. There is an ambiguity in the sign of the this phase, however it does not effect the final conclusion.}

\begin{figure}
\begin{center}
\includegraphics[scale=0.42]{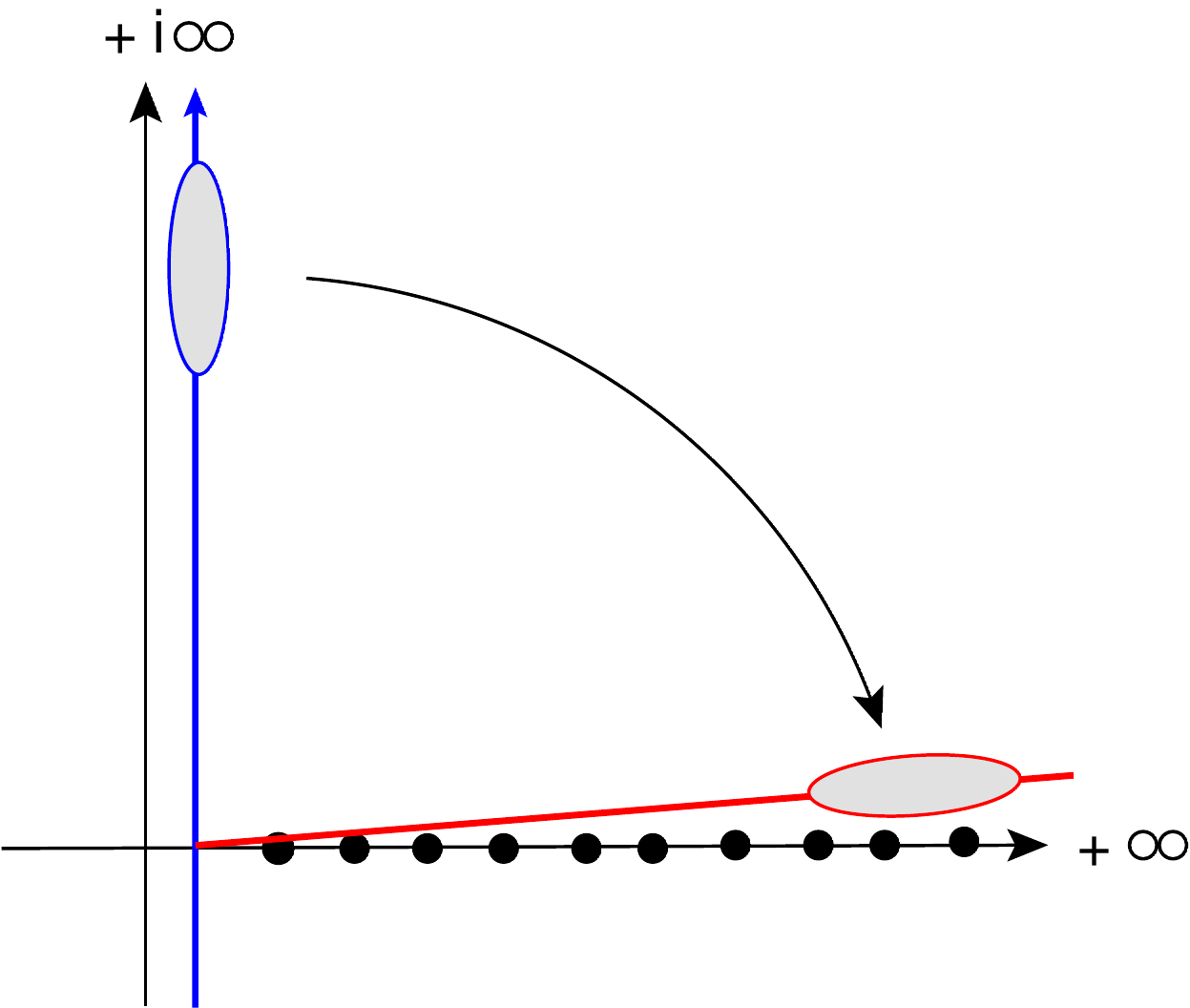}
\caption{The contour for the Mellin integral is defined along the imaginary axes with a small real positive part. We are taking the asymptotic value at $s\rightarrow i\infty$ and analytically continuing it near the positive real axes, which we are allowed to do so since all the poles of the integrand lies on the real axes.  }
\label{fig1}
\end{center}
\end{figure}

Now on top of the flat space limit we consider the limit where $|s|>>|t|\sim\Delta>>1$,
\eqa\label{Def2}
\mathcal{A}(u,v)&\approx&-{1\over 4\sqrt{2\pi}} e^{-\Delta}\int^{i\infty}dtds\widetilde{M}(s,t)\frac{u^{\frac{t}{2}}}{v^{\frac{s}{2}-\Delta}e^{i\pi s}}\left(\frac{s}{2}\right)^{\frac{t}{2}}\nonumber\\
&&(-1)^{\frac{s}{2}-\Delta}\,\left(\Delta-\frac{t}{2}\right)^{\Delta-\frac{t}{2}}\,\nonumber\\
&=&-\frac{1}{4\sqrt{2\pi}}\left(\frac{{-}v}{e}\right)^{\Delta}\int^{i\infty}dtds\widetilde{M}(s,t)e^{\frac{t}{2}\log\left(u\frac{s}{2\Delta{-}t}\right)}\nonumber\\
&&\,e^{-\frac{s}{2}\log{v}}\,e^{\Delta\log(\Delta-t/2)}e^{i\pi\frac{s}{2}}\,,
\eqae
Recall that the Mellin integral is along the imaginary axe, hence due to the factor $e^{-\frac{s}{2}\log v}$, the large $s$ region is generically highly oscillating and thus do not contribute. However in the limit where $v\rightarrow1$, such that 
\eq\label{xlimit}
s\log v<<1\,,
\eqe
the oscillation is supressed and the large $s$ limit survives. In similar fashion, the factor $e^{\frac{t}{2}\log\left(u\frac{s}{2\Delta-t}\right)}$ is highly oscillating for large $t$, unless
\eqa
u\sim \frac{2\Delta-t}{s}\,.
\eqae 
Since the relative scales we are considering is $|s|>>|t|\sim\Delta>>1$, this implies that $u$ is near to zero in order for this limit to survive. This establishes our claim that taking $v\rightarrow 1, u\rightarrow 0, \Delta>>1$ for the reduced correlator probes the asymptotic Regge limit
\eq 
|s|>>|t|>>1,
\eqe 
for a flat-space S-matrix. By analytically continuing to the real axes as in fig.\ref{fig1}, this is precisely what was studied by~\cite{Caron-Huot:2016icg}.

Using the fact that the flat-space limit of the Mellin amplitude correspond to a flat-space S-matrix, where it's unitarity and requisite permutation symmetry is inherited from the unitarity and crossing symmetry of the CFT correlator, in the Regge-limit we can substitute the Mellin amplitude with the general form,
\eq
M(s,t)=F(t) s^{j(t)}\,.
\eqe
Generically the above form is valid for $|s|>>|t|>1$. Here we assume that the same form persists in the asymptotic Regge limit. Note that it can also be argued from taking the flat space limit of the Regge-limit Mellin amplitude studied in~\cite{CRegge}. Substituting this for $M(s,t)$ in eq.\eqref{Def2} we find, 
\eqa
\mathcal{A}(u,v)&\approx &-\frac{1}{4\sqrt{2\pi}}\left(\frac{-v}{e}\right)^{\Delta}\nonumber\\&&\int^{\infty}ds dt\,F(t) e^{-h(s,t)}e^{g(t,u,v)}\,.
\eqae
where $h(s,t)={s\over 2}\ln(v)-(\frac{t}{2}+j(t))\ln s$. 

Now we rotate the integration contour to slightly off the real axes as shown in fig.\ref{fig1}, which is allowed since all the poles lie on the real axes. It is slightly off to avoid hitting those poles. Since we assumed that the Mellin amplitude satisfy eq.(\ref{ReggeS}), this implies that there is a saddle point in the $s$ integral for some value of $t$. The saddle point of the integral $s$ near the large $s$ regime is determined by:
\eq
{1\over 2}\ln(v)-(\frac{t}{2}+j(t))\frac{1}{s}=0\,.
\eqe
Recall that we are in the regime where eq.(\ref{xlimit}) holds, and thus the saddle point is identified as:
\eq\label{i}
\frac{t}{2}+j(t)<<1\,.
\eqe  
Now assume that eq.(\ref{ReggeS}) holds for some value of $t$ that satisfies $|t|>>1$, this in turn requires eq.(\ref{i}) to also hold in this regime. This immediately bound the highest degree that $j(t)$ can have to be 1, i.e. at large $|t|$ the Regge trajectory must be linear :
\eq
j(t)\sim t.
\eqe
Thus the fact that the Mellin amplitude has vanishing Regge asymptotic implies that the Regge trajectory must be linear in the asymptotic Regge limit, $|s|>>|t|>>1$.  

We can ask where precisely is the saddle point. Let $j(t)=at$ where $a$ is a dimensionful constant. \newline Since $j(t)=n$ with $n\in \mathbb{N}$ correspond to the mass of the particle, this imply that $a>0$, and hence the saddle point occurs when $t$ is negative.   
\section{Discussions and Conclusions}
In this note, we utilise the intriguing relation between Mellin amplitudes and flat space S-matrix, to study general properties of the latter with vanishing Regge asymptotic. We consider the four-point correlation function of scalar operators $\mathcal{O}$. For CFTs whose composite operator ${\footnotesize \mathcal{O}(\overleftrightarrow{\partial})^{2n}\mathcal{O}}$ acquires an anomalous dimension that is perturbative in some parameters, the Mellin amplitude has a separate piece that can be considered as a tree-level scattering amplitude. Taking the flat-space limit, which correspond to large scaling dimensions, the Mellin amplitude can be identified as a flat space S-matrix. The requisite properties such as unitarity and permutation invariance, are then directly inherited from the unitarity and crossing symmetry of the correlation function. 

We consider CFTs whose Mellin amplitude vanish in the Regge limit for particular values of $t$. CFTs that satisfy this criteria can be identified through the study of three-point correlation functions involving two $\mathcal{O}$ and one ${\footnotesize \mathcal{O}(\overleftrightarrow{\partial})^{2n}\mathcal{O}}$, where the spin of the latter is taken to be large. Matching with free field theory we indeed find regions for vanishing Regge behaviour. In the usual AdS/CFT duality, this corresponds to the small t'Hooft limit. 

The vanishing of Mellin amplitude for large $s$ implies the existence of a saddle point in the Mellin integral. To study this limit of the Mellin amplitude, we show that by taking $(v\rightarrow 1, u\rightarrow 0, \Delta>>1)$, corresponding to the asymptotic Regge limit, $|s|>>|t|>>1$, of the flat space S-matrix. For a saddle point to exist, we show that the Regge trajectory must be linear. Thus for CFTs whose Mellin amplitude has vanishing Regge asymptotics, the corresponding flat space limit must become string like in the asymptotic Regge limit, in agreement with~\cite{Caron-Huot:2016icg}. 

Note that while this result was derived for the simplest scalar interactions, we expect similar conclusion for the scattering of higher spin states. The reason is that this result was dimension independent. Thus if one considers the scattering of massive higher spin states with the same asymptotic Regge behaviour, upon dimensional reduction one must obtain massive scalar amplitudes with the same asymptotic behaviour. Thus one would expect that the higher spin S-matrix has the same linear Regge trajectory.
 
 \vspace{5mm}


\vspace{-0.4cm}

\section{Acknowledgements}

We would like to thank Luis Alday, Gregory Korchemsky,  Zohar Komargodski and Shiraz Minwalla for the discussions. Y-t Huang is supported by MOST
under the grant No. 103-2112-M-002-025-MY3. The work of C.C. is supported in part by the National Center for Theoretical Science (NCTS), Taiwan.
\end{document}